\documentclass[zpreprint,zbstpl]{./paper}
\usepackage[american]{babel}
\usepackage{graphpap}
\begin{document}

\title{
Stop production in R-parity \\
violating supersymmetry at Tevatron
}                                                       
                    
\author{L. Bellagamba\\
Istituto Nazionale di Fisica Nucleare\\ Via Irnerio 46, I-40126 Bologna, 
Italy\\
lorenzo.bellagamba@bo.infn.it
}
\date{ }

\abstract{ 
Constraints for stop production in R-parity violating 
supersymmetry at Tevatron have been evaluated using the D0 limits 
for first generation leptoquarks.
Such limits have been converted in constraints for the 
$\lambda'_{131}$ R-parity violating coupling as a function of the stop
mass for different Minimal Supersymmetric Standard Model (MSSM) scenarios 
and compared with Atomic Parity Violation and HERA results.
The D0 limits have also been interpreted in terms of constraints
on the parameters of the minimal Supergravity model.
For a part of the considered MSSM parameter space and for stop
mass less than 240 GeV, Tevatron results are the best to date. 
}

\makezeustitle
\pagestyle{plain}
%
\section{Introduction}
New exotic states connecting lepton and quark sectors  
naturally arise in grand unification theories~\cite{gut} that arrange
quarks and leptons in common multiplets (leptoquarks, LQ) or in 
supersymmetric (SUSY) models
that violate $R$-parity~\cite{rp} (squarks, $\tilde{q}$). These states 
should be
produced by the same processes at the colliders
but differ for the decay modes.
Leptoquarks exclusively
decay via a Yukawa coupling in a lepton and a quark while squarks 
have further gauge decay modes in a gaugino and a quark or in a
squark of different flavor and a $W$. In proton-anti-proton collisions at
Tevatron both leptoquarks and squarks should be pair produced via ordinary
gauge couplings dominantly by quark-anti-quark annihilation and 
gluon-gluon fusion. A detailed description of stop phenomenology in 
$R$-parity violating SUSY at Tevatron can be found in~\cite{RPVTevee}.\\
The study presented here focuses on the stop\footnote{The SUSY partners of
the left- and right-handed top, $\tilde{t}_L$ and $\tilde{t}_R$, mix
together in  two mass eigenstates,  $\tilde{t}_1$ and $\tilde{t}_2$,
which  are  strongly non-degenerate  due  to  the  large top  mass.  
Throughout the paper, with the symbol $\tilde{t}$ we always refer to the 
lightest stop, $\tilde{t}_1$.} 
($\tilde{t}$, supersymmetric 
partner of the top-quark) production, since it is predicted to be the 
lightest sfermion in a large variety of possible supersymmetric scenarios.
In $R$-parity violating SUSY a Yukawa coupling 
$\lambda'_{131}$ (where the subscripts are generation indices) rules
the partial width of the reaction
$\tilde{t} \rightarrow e q$\footnote{Here and in the following
we generically refer to both stop and anti-stop; $e$ 
hence denotes both a positron or an electron and $q$ a $d$- or an 
anti-$d$-quark. Throughout the paper we also use "electron" to denote
both an electron or a positron.}
while the partial width of the
gauge decays depend on the SUSY scenario considered.\\
In this paper we evaluate $95\%$ confidence level (CL) limits for 
the stop in $R$-parity violating SUSY at Tevatron, using 
recent D0 limits\footnote{The D0 limits obtained assuming 
that both LQs decay to $eq$ (final states with two electrons and two jets)
have been considered in this study. The combined limits, obtained looking
also at the $\nu q$ decay of one of the two LQs (final states with one 
electron, two jets and missing transverse energy), cannot be applied to
the stop.}   
~\cite{rpbrD0} (table~\ref{tab-br}) on the mass of first 
generation 
leptoquarks\footnote{First 
generation leptoquarks are assumed to couple only to quarks and leptons of
the first family.}
as a function of the branching ratio to an electron 
and a quark. Such results have been 
obtained using Tevatron run II data corresponding 
to an integrated luminosity of $\simeq 250$ pb$^{-1}$.  
Similar limits have been previously evaluated using 
$\simeq 120$ pb$^{-1}$ of integrated luminosity collected
during the Tevatron run I~\cite{RPVTevee}.\\
\begin{table}[htb]
\begin{center}
\begin{tabular}{ l  r  r  r  r  r  r } \hline \hline
 Br($LQ \rightarrow eq$) & 0.5 & 0.6 & 0.7 & 0.8 & 0.9 & 1.  \\
\hline
 LQ mass limit (GeV) & 158 & 180 & 203 & 220 & 232 & 241 \\
\hline \hline
\end{tabular}
\caption{
D0 $95\%$ CL exclusion limits for first
generation leptoquark mass as a function of the branching ratio to
an electron and a quark.}
\label{tab-br}
\end{center}
\end{table}
Constraints on stop production in $R$-parity violating SUSY have 
been already reported by the H1~\cite{H1:SUSY} and ZEUS~\cite{ZEUS:SUSY} 
collaborations at HERA, where the stop can be resonantly produced by
electron-quark fusion via the $\lambda'_{131}$ coupling. Indirect limits also come 
from low-energy experiment on atomic parity violation (APV)~\cite{APV, APV2}.
\section{Results}
\begin{figure}[htb]
\begin{minipage}[t]{3.5cm}
\epsfxsize=8cm   
\epsfbox{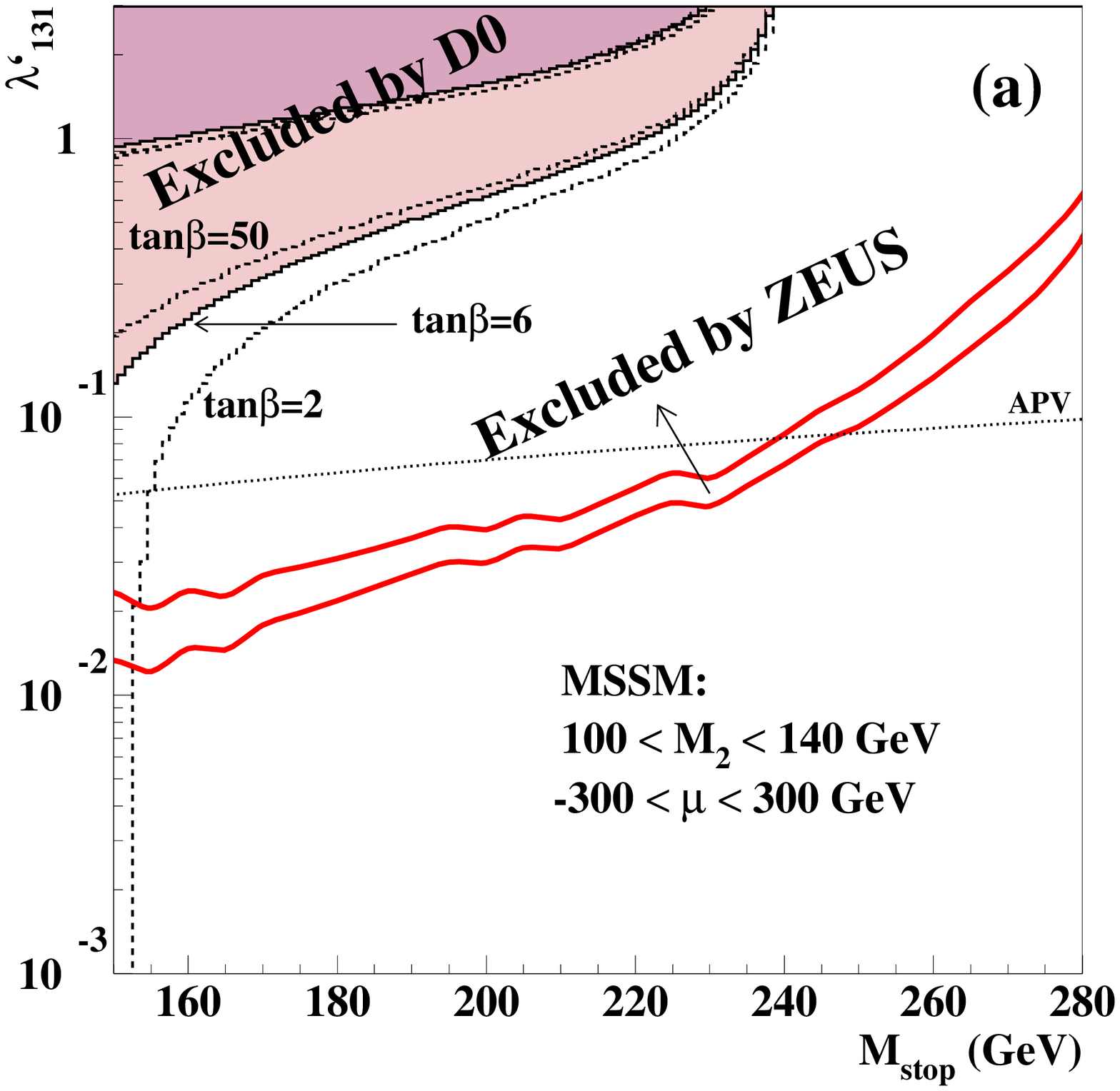}
\end{minipage} \hfil
\begin{minipage}[t]{3.5cm}
\epsfxsize=8cm   
\epsfbox{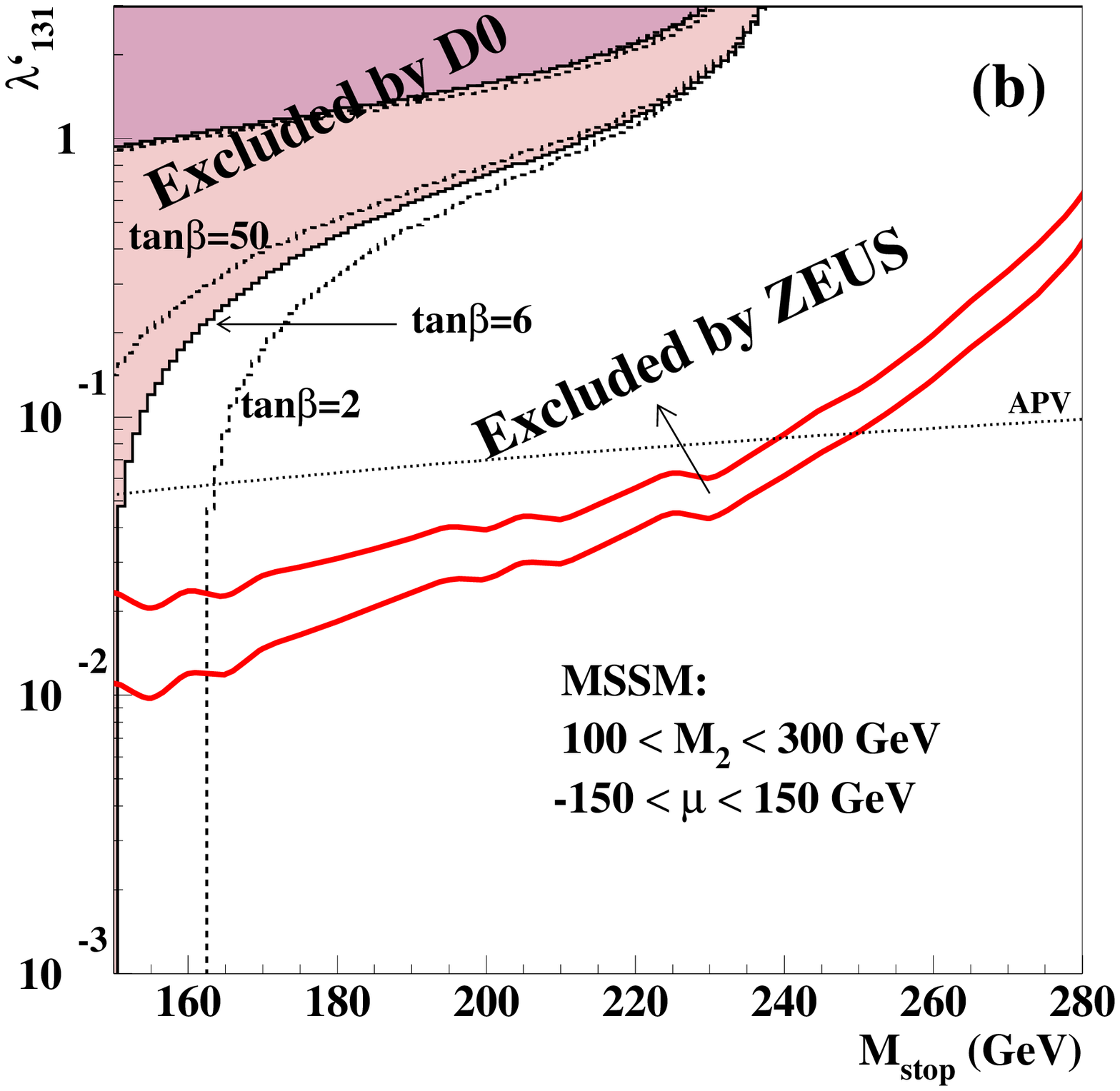}
\end{minipage} \hfil
\caption{$95\%$ CL exclusion limits for $\lambda'_{131}$ as a function of 
the stop mass for low $M_2$ (a) and low $|\mu|$ (b) MSSM
scenarios. The lighter (darker) area denotes the region excluded by
D0 for part of (all) the considered scenarios for $\tan\beta=6$.
The thick solid lines show the ZEUS limits for $\tan\beta=6$; the region 
between 
the two lines (above the upper line) is excluded for part of (all) the 
considered scenarios. D0 results for $\tan\beta=2$ and $50$ (dashed lines)
are also reported. The dotted line is the indirect constraint from
APV measurements.}
\label{fig-mssm1}
\end{figure}

\begin{figure}[htb]
\begin{center}
\epsfig{figure=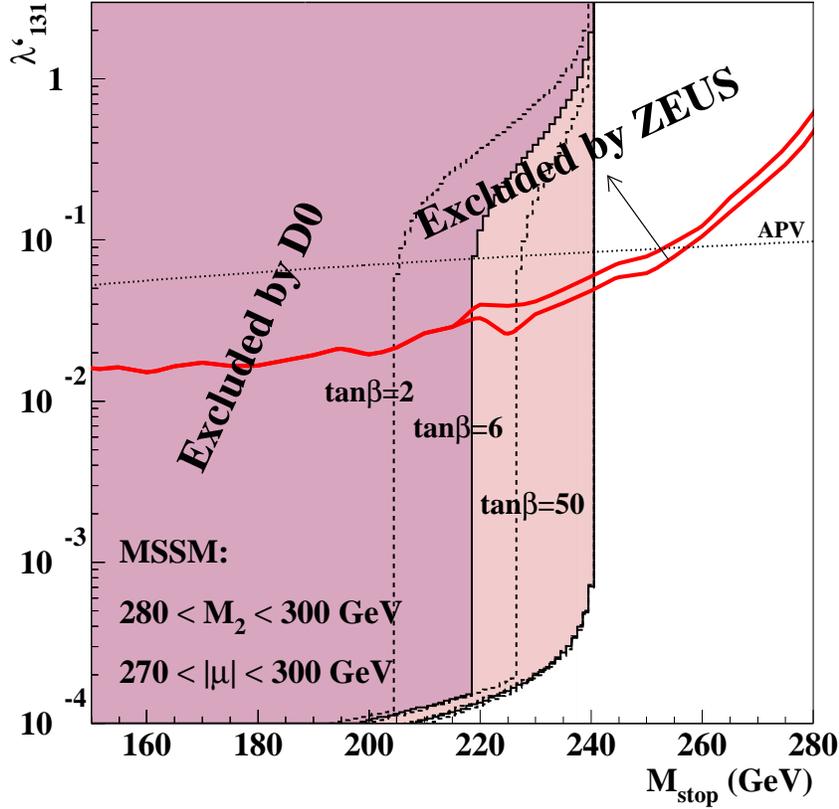,height=12cm}
\caption{$95\%$ CL exclusion limits for $\lambda'_{131}$ as a function of 
the stop mass for high $M_2$ and high $|\mu|$ MSSM scenarios. The lighter 
(darker) area denotes the region excluded by
D0 for part of (all) the considered scenarios for $\tan\beta=6$.
The thick solid lines show the ZEUS limits for $\tan\beta=6$: the region 
between 
the two lines (above the upper line) is excluded for part of (all) the 
considered scenarios. D0 results for $\tan\beta=2$ and $50$ (dashed lines)
are also reported. The dotted line is the indirect constraint from
APV measurements.}
\end{center}
\label{fig-mssm2}
\end{figure}

\begin{figure}[htb]
\begin{minipage}[t]{3.5cm}
\epsfxsize=8cm   
\epsfbox{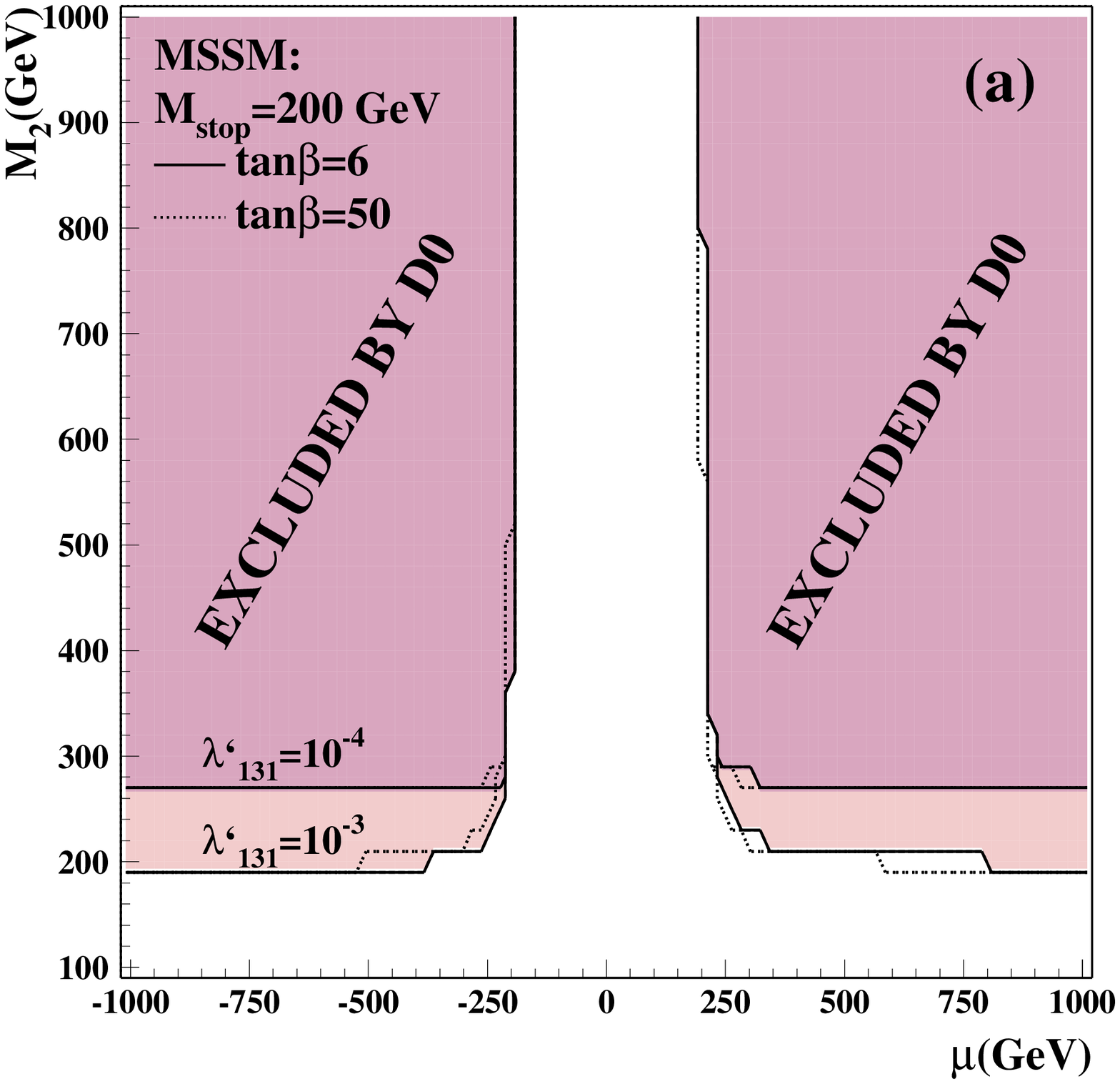}
\end{minipage} \hfil
\begin{minipage}[t]{3.5cm}
\epsfxsize=8cm   
\epsfbox{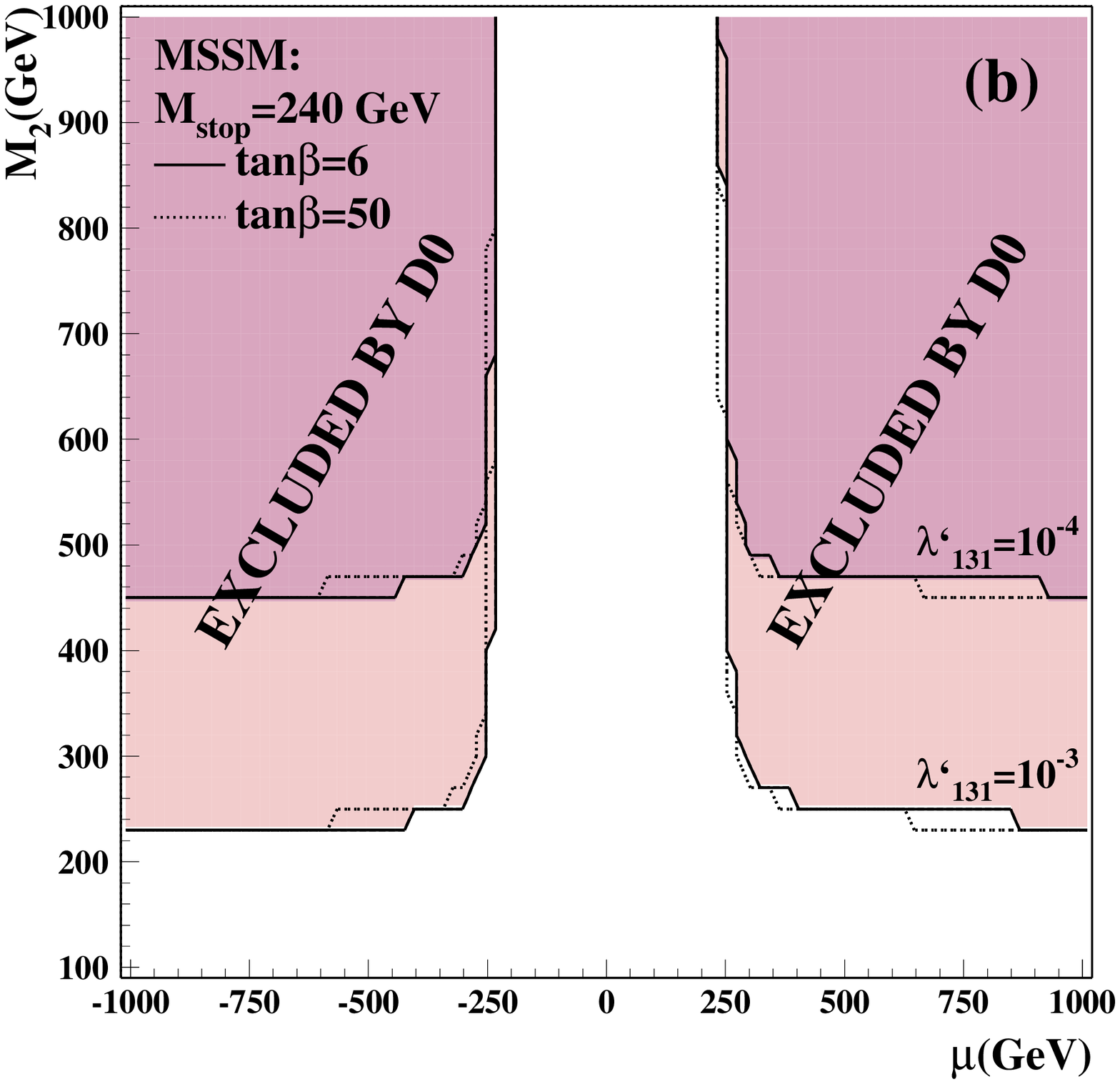}
\end{minipage} \hfil
\caption{$95\%$ CL D0 exclusion limits in the $\mu$-$M_2$ plane
for MSSM scenarios with $\tan\beta = 6$ in the case of $M_{stop}=200$ (a) 
and $240$ GeV (b). The darker area denotes the region excluded for a coupling 
$\lambda'_{131}=10^{-4}$, the lighter area is the further exclusion
region for $\lambda'_{131}=10^{-3}$. 
Limits for $\tan\beta = 50$ (dashed lines) are also reported.}
\label{fig-largescan}
\end{figure}

The constraints on stop production have been obtained using the program
SUSYGEN 3~\cite{SUSYGEN1, SUSYGEN2} to describe different SUSY scenarios. 
The only $R$-parity violating Yukawa coupling different from 
zero was assumed to be $\lambda'_{131}$.
A particular scenario is ruled out at $95\%$ CL if it predicts for the 
stop a branching ratio to $eq$ already excluded by the D0 leptoquark
analysis.\\
An unconstrained Minimal Supersymmetric Standard Model (MSSM)
scenario was initially considered.
The masses of the sfermions are free parameters of the model while the  
masses  of  the  neutralinos, charginos  and  gluinos  are
determined by the following parameters: the mass term $\mu$ which
mixes the Higgs superfields, the soft SUSY-breaking parameters $M_1$,
$M_2$  and  $M_3$  for  the  $U(1)$,  $SU(2)$  and  $SU(3)$
gauginos\footnote{The gaugino masses are assumed to converge to a 
common mass $m_{1/2}$ at the GUT scale, leading to relations among
$M_1$, $M_2$  and  $M_3$. The SUSY scenario is thus defined by just one
of the three soft SUSY-breaking masses.},
respectively, and  $\tan{\beta}$, the ratio of  the vacuum expectation
value of the two neutral scalar Higgs fields. 
The mass of the lightest stop was varied in the range $150-280$ GeV 
while the masses of the other squarks and all the sleptons have been set 
at 1 TeV. In order to have a direct comparison between HERA and Tevatron 
sensitivities, we considered the same scenarios investigated by the
ZEUS collaboration which published a search for stop production
looking both at the $eq$ and $b \chi^+_1$ (where $\chi^+_1$ is the 
lightest chargino) decays~\cite{ZEUS:SUSY}: 
$\tan{\beta}=6$, 
$-300 \le \mu \le 300$ GeV and $100 \le M_2 \le 300$ 
GeV. Scenarios leading to a neutralino mass less than 
$30$ GeV, already excluded by LEP results~\cite{epj:c19:397}, were 
discarded.
Figures~\ref{fig-mssm1}
and~\ref{fig-mssm2} show limits on $\lambda'_{131}$ 
as a function of the stop mass for three choices 
of the parameter space. 
The lighter (darker) area denotes the region excluded by
D0 for part of (all) the considered scenarios. 
The thick solid lines show the ZEUS limits; the region 
between 
the two lines (above the upper line) is excluded for part of (all) the 
considered scenarios. H1 published similar results~\cite{H1:SUSY}.
Both D0 and ZEUS limits have a weak dependence of $\tan\beta$;
D0 results for $\tan\beta=2$ and $50$ (dashed lines) are also reported. 
The dotted line is the indirect constraint from
atomic parity violation (APV)~\cite{APV, APV2} measurements, the 
region above the line is excluded.\\
For low values of $M_2$ (fig.~\ref{fig-mssm1} a) 
or $|\mu|$ (fig.~\ref{fig-mssm1} b) D0 constraints are
not competitive with ZEUS ones for stop masses larger than $\sim 150$ GeV. 
Indeed, in this region of the parameter space, the first chargino is 
lighter than the stop and the gauge stop decay 
$\tilde{t} \rightarrow b \chi_1^+$ dominates over the $\tilde{t} 
\rightarrow e q$ decay.\\
Quite different is the high $M_2$ - high $|\mu|$ case
(fig.~\ref{fig-mssm2}) 
where, due to the high chargino mass, the stop 
branching ratio to $eq$ is dominant. Since at Tevatron, contrary to
HERA, the stop production cross section is independent of the 
$\lambda'_{131}$ coupling, much lower coupling values can be probed in
this case. It is interesting to note that, in this range of parameters,
the excluded ZEUS region is almost completely
superseded by D0 and APV limits.\\
In order to test the validity of the D0 constraints
towards higher $|\mu|$ and $M_2$, a larger scan of the SUSY parameter 
space
was performed for fixed stop mass 
and for two values of $\lambda'_{131}=10^{-3}$ and $10^{-4}$.
Fig.~\ref{fig-largescan} a and b show the D0 exclusion limits for
$\tan\beta = 6$ in the 
$\mu$-$M_2$ plane in the case of $M_{stop}=200$ and $240$ GeV, 
respectively. The darker area denotes the region excluded for a coupling 
$\lambda'_{131}=10^{-4}$, the lighter area is the further exclusion
region for $\lambda'_{131}=10^{-3}$. 
The D0 constraints are still valid at the highest $|\mu|$ and $M_2$ 
values and show negligible dependence of $\tan\beta$ as shown by the 
limits for $\tan\beta = 50$ (dashed lines) which are also reported. The 
limits have also a small dependence
of the slepton masses, the scan was repeated lowering the mass of 
all sleptons from $1$ TeV to $100$ GeV, as expected no remarkable 
variations were observed.\\
Even if we focused on the coupling $\lambda'_{131}$ in order to have a
direct comparison with HERA constraints, Tevatron limits are largely 
independent of the flavor of the quark coming from stop decay. The 
limits hence apply to the generic coupling $\lambda'_{13j}$
with $j=1,2,3$.\\
Respect to the previously reported Tevatron limits on stop in 
$R$-parity violating SUSY~\cite{RPVTevee} our results are
based on a twofold integrated luminosity, further
a larger scan of MSSM parameter space has been performed.\\ 
Besides $\lambda'_{13j}$ also $R$-parity violating MSSM scenarios 
involving the coupling $\lambda'_{23j}$ have been 
investigated using Tevatron Run II dimuon data~\cite{RPVTevmumu}.\\
The study has been also extended to the minimal Supergravity model
(mSUGRA)~\cite{mSUGRA1, mSUGRA2, mSUGRA3}. In this model the number
of free parameters is further reduced by assuming, 
beside the common gaugino mass $m_{1/2}$, also a common sfermion
mass $m_0$ at the GUT scale. Radiative corrections are assumed to drive
the electroweak symmetry breaking (REWSB), leading to consistency 
relations that allow the complete model to be fixed by only five 
parameters: $m_0$, $m_{1/2}$, $\tan{\beta}$, the sign of $\mu$ and the
common trilinear coupling $A_0$.
The mSUGRA scenario was modelled by SUSYGEN 3~\cite{SUSYGEN1, SUSYGEN2} 
which makes use of the SUSPECT 2.1~\cite{SUSPECT} program to solve the
REWSB consistency relations that determine the sparticle masses at the
electroweak scale.
The same scenarios analysed by the ZEUS~\cite{ZEUS:SUSY} 
collaborations ($\tan{\beta}=6$, 
$\lambda'_{131}=0.3$, $A_0=0$, $m_0 \leq 300$ GeV, $m_{1/2}\leq 180$ GeV) 
were studied. In this case Tevatron limits are not
competitive with HERA results, since, being the chargino always lighter 
than the stop, the gauge decays of the
stop are dominant for all the scenarios considered.
%
%
\section{Conclusions}
Limits at $95\%$ CL for stop production in $R$-parity 
violating SUSY at Tevatron have been evaluated using D0 constraints for 
first generation leptoquarks obtained using data collected in run II 
and corresponding to an integrated luminosity of 
$\simeq 250$ pb$^{-1}$.\\ 
Both unconstrained MSSM and mSUGRA scenarios have been considered.
For the considered mSUGRA scenarios ($\tan{\beta}=6$, 
$\lambda'_{131}=0.3$, $A_0=0$, $m_0 \leq 300$ GeV, $m_{1/2}\leq 180$ GeV),
Tevatron limits are weaker than HERA ones.
In the MSSM case, for large $M_2$ ($280 < M_2 < 300$ GeV) and $|\mu|$ 
($270 < |\mu| < 300$ GeV), 
Tevatron results, for stop masses less than 240 GeV, are the best to date. 
The Tevatron limits remain competitive also for higher $|\mu|$ and $M_2$
and have a negligible dependence of $\tan\beta$ and slepton masses.
For lower values of $M_2$ or $|\mu|$ HERA limits are stronger.
\section{Acknowledgements}
We would like to thank M.~Corradi, E.~Gallo and M.~Kuze for critical 
reading of the manuscript and helpful comments.
\bibliographystyle{plain}	
\bibliography{stop_tevatron}

\providecommand{\etal}{et al.\xspace}
\providecommand{\coll}{Coll.\xspace}
\begin{thebibliography}{10}
\bibitem{gut}
J.~C.~Pati and A.~Salam, Phys. Rev. {\bf D10} (1974) 275;\\
H.~Georgi and S.~L.~Glashow, Phys. Rev. Lett. {\bf 32} (1974) 438;\\
P.~Langacker Phys. Rep. {\bf 72} (1981) 185
\relax
\bibitem{rp}
S.~Weinberg, Phys. Rev. {\bf D26} (1982) 287;\\
N.~Sakai and T.~Yanagida, Nucl. Phys. {\bf B197} (1982) 83;\\
S.~Raby and F.~Wilczek Phys. Lett. {\bf B212} (1982) 133 
\relax
\bibitem{RPVTevee} 
S.~Chakrabarti, M.~Guchait and N.~K.~Mondal, Phys. Rev. D 
{\bf 68} (2003) 015005, hep-ph/0301248
\relax
\bibitem{rpbrD0}
D0 Collab., V.~M.~Abazov \etal, Phys. Rev. D Rapid Comm.
{\bf 71} (2005) 071104, hep-ex/0412029
\relax
\bibitem{H1:SUSY}
H1 Collab., A.~Aktas \etal, Eur. Phys. J. {\bf C36} (2004) 425, 
hep-ex/0403027
\relax
\bibitem{ZEUS:SUSY}
ZEUS Collab., S.~Chekanov \etal, DESY-06-144, submitted to Eur. Phys. J.,
hep-ex/0611018
\relax
\bibitem{APV}
H.K.~Dreiner, 
'Perspectives on Supersymmetry', Ed. by G.L. Kane, World Scientific,
hep-ph/9707435
\relax
\bibitem{APV2}
P.~Langacker, Phys. Lett. {\bf B256} (1991) 277
\relax
\bibitem{SUSYGEN1}
S.~Katzanevas and P.~Morawitz, Comput. Phys. Commun. {\bf 112} 
(1998) 227, hep-ph/9711417
\relax 
\bibitem{SUSYGEN2}
N.~Ghodbane, S.~Katzanevas, P.~Morawitz and E.~Perez, SUSYGEN 3, 
hep-ph/9909499
\relax 
\bibitem{epj:c19:397}
L3 Collab., M.~Acciarri \etal, Eur. Phys. J. {\bf C19} (2001) 397, 
hep-ex/0011087
\relax
\bibitem{RPVTevmumu} 
S.~Chakrabarti, M.~Guchait and N.~K.~Mondal, Phys. Lett. 
{\bf B600} (2004) 231, hep-ph/0404261
\relax
\bibitem{mSUGRA1}
M.~Drees and M.~M.~Nojiri, Nucl. Phys. {\bf B369} (1992) 54
\relax
\bibitem{mSUGRA2}
H.~Baer and X.~Tata, Phys. Rev. {\bf D47} (1993) 2739
\relax
\bibitem{mSUGRA3}
G.~L.~Kane, C.~Kolda, L.~Roszkowski and J.~D.~Wells, Phys. Rev. {\bf 
D49} (1994) 6173, hep-ph/9312272
\relax
\bibitem{SUSPECT}
SUSPECT 2.1 code written by A.~Djouadi, J.~L.~Kneur and G.~Moultaka,
hep-ph/0211331
\relax 
\end{thebibliography}
\end{document}